\documentclass[aps,prl,twocolumn,letterpaper,superscriptaddress,showpacs]{revtex4}
\usepackage{CJK}
\usepackage{keyval}%
\usepackage{graphicx}
\usepackage{dcolumn}
\usepackage{bm}
\usepackage{color}

\setkeys{Gin}{width=0.8\columnwidth,clip=true}
\newcommand{\ignore}[1]{}

\begin{document}
\begin{CJK*}{UTF8}{bsmi}
\title{Unfolding first-principles band structures}
\author{Wei Ku (%
顧威
)}
\affiliation{Condensed Matter Physics and Materials Science
Department, Brookhaven National Laboratory, Upton, New York 11973,
USA}
\affiliation{Physics Department, State University of New York, Stony
Brook, New York 11790, USA}
\author{Tom Berlijn}
\affiliation{Condensed Matter Physics and Materials Science
Department, Brookhaven National Laboratory, Upton, New York 11973,
USA}
\affiliation{Physics Department, State University of New York, Stony
Brook, New York 11790, USA}

\author{Chi-Cheng Lee (%
李啟正
)}
\affiliation{Condensed Matter Physics and Materials Science
Department, Brookhaven National Laboratory, Upton, New York 11973,
USA}%

\date{\today}

\begin{abstract}
A general method is presented to unfold band structures of first-principles super-cell
calculations with proper spectral weight, allowing easier visualization of the electronic
structure and the degree of broken translational symmetry.
The resulting unfolded band structures contain additional rich information from the Kohn-Sham
orbitals, and absorb the structure factor that makes them ideal for a direct comparison
with angular resolved photoemission spectroscopy experiments.
With negligible computational expense via the use of Wannier functions, this simple method has
great practical value in the studies of a wide range of materials containing impurities,
vacancies, lattice distortions, or spontaneous long-range orders.
\end{abstract}


\maketitle
\end{CJK*}

The electronic band structure is no doubt one of the most widely applied analysis tools in the
first-principles electronic structure calculations of crystals, especially within the Kohn-Sham
framework \cite{KohnSham} of density functional theory~\cite{HohenbergKohn}.
It contains the basic ingredients to almost all the textbook descriptions of 
crystal properties (e.g. transport, optical and magnetic properties, and the semiclassical
treatment \cite{AshcroftMermin}).
Furthermore, the theoretical band structure, when formulated within the quasi-particle picture of
the one-particle Green function, has a direct experimental connection with angular-resolved
photoemission spectroscopy (ARPES).

\begin{figure}[tbp]
\includegraphics[width=0.9\columnwidth,clip=true]{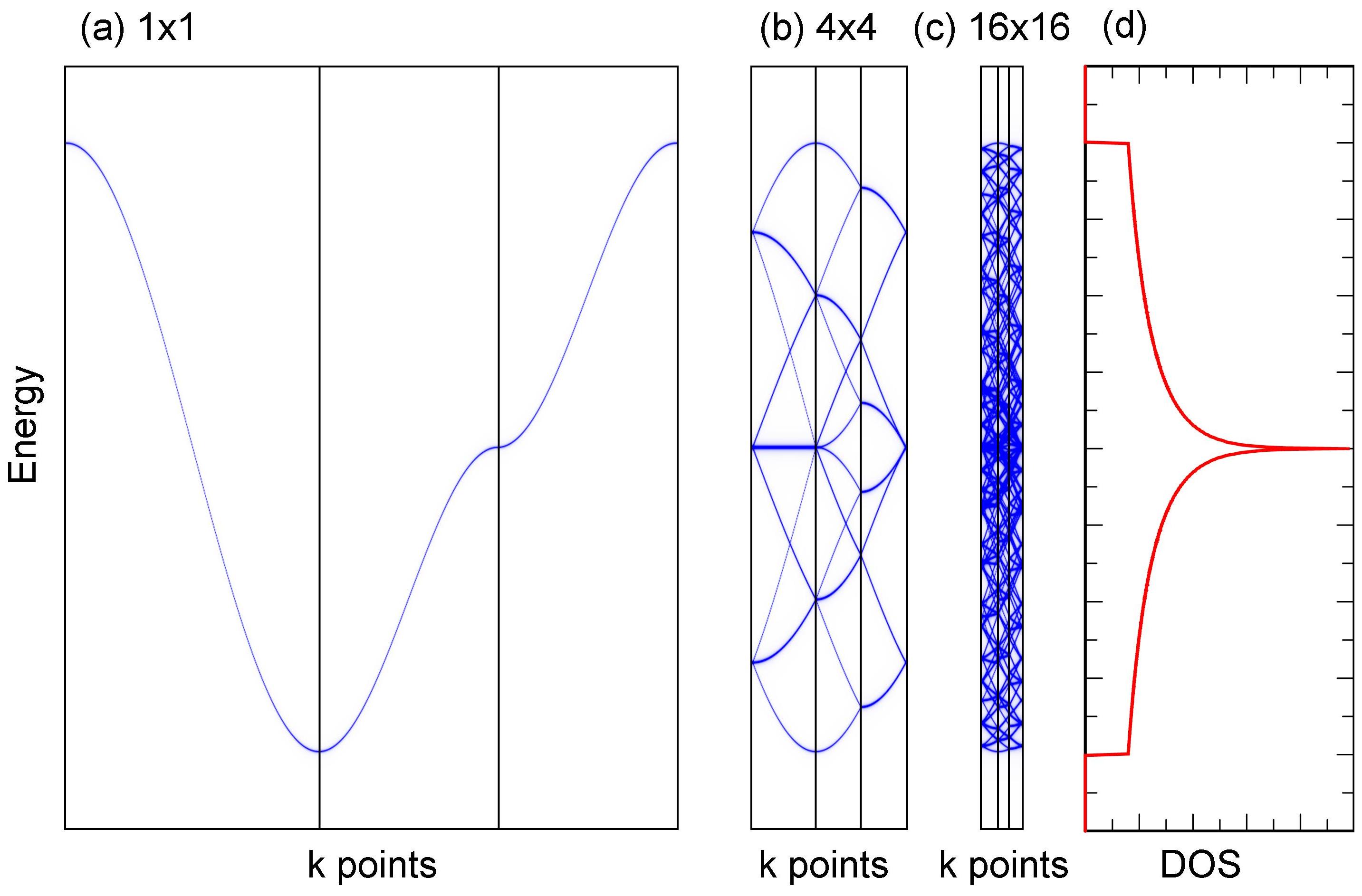}
\caption{\label{fig:fig1} (color online)
Illustration of band folding in the super cell calculations:
(a) band structure of a 2D one-band first-neighbor tight-binding model, (b) the same obtained
from a 4x4 super-cell calculation, and (c) the same obtained from a 16x16 super-cell calculation.
Pannel (d) shows the DOS.
}
\end{figure}

However, the usefulness of the band structure, as well as the agreement with ARPES spectra,
diminishes rapidly when a large ``super cell'' is involved.
The use of super cells is a common practice in modern first-principles studies when the original
periodicity of the system is modified via the introduction of ``external'' influences from
impurities or lattice distortions.
They are also widely applied in the presence of spontaneous translational symmetry breaking,
say by a charge density wave, a spin density wave, or an orbital ordering.
As illustrated in Fig.~\ref{fig:fig1}, when the period of the super cell grows longer, the
corresponding first Brillouin zone of the super cell (SBZ) shrinks its size.
In turn, bands in the first Brillouin zone of the normal cell (NBZ) get ``folded'' into the SBZ.
For a very large super cell, the resulting SBZ can be tiny in size but contain a large number of
``horizontal'' looking bands that no longer resemble the original band structure or the
experimental ARPES spectra, and cease to be informative besides giving a rough visualization of
the density of states (DOS).
The information is now hidden in the Kohn-Sham orbitals, instead of the dispersion of the bands.

In this Letter, by explictly utilizing these Kohn-Sham orbitals, we present a method to unfold
the band structure of the SBZ back to the larger NBZ with proper spectral weight.
Making use of the corresponding Wannier functions, the method can be greatly simplified to
negligible computational cost.
The resulting unfolded band structure incorporates explicitly the structure factor and thus
facilitates significantly a direct comparison with ARPES experiments.
Furthermore, the unfolded band structure illustrates very clearly the influence of the symmetry
breaker (e.g.: impurities, vacancies, dopants, lattice distortions) via direct comparison with the
nominal normal-cell band structure.
In the case of spontaneous symmetry breaking, it gives a direct visualization of the strength of
each band's coupling to the order parameters.
In light of the amazingly rich information, we expect countless applications of this simple
method to a wide range of studies employing super cells, including systems with charge density
wave, spin density wave, or orbital ordering, and in the studies of impurities and lattice
distortions, to name a few.

Theoretically, the folding of the bands results from the introduction of additional coupling,
$V_{kj,k\prime j\prime}$, between the originally uncoupled Kohn-Sham orbitals $|kj\rangle$ and
$|k\prime j\prime\rangle$ in the NBZ.
(Here $k$ and $j$ denote the crystal momentum and the band index.)
This coupling extends the period of Kohn-Sham orbitals to a longer one compatible with the
size of the super cell.
Equivelantly, this coupling, no matter how small it is, mixes the original orbitals of different
normal-cell crystal momentum k and forces us to label them with a supercell crystal momentum K
as the new quantum number in the SBZ.
(In the following, upper-/lower-case symbols refer to variables corresponding to the super/normal
cell, respectively.)
Our method is based on the simple idea that unless $V$ is extremely strong, it is much more
convenient and informative to represent the band structure or more precisely the spectral
function $A = - {\rm{Im}} G / \pi$ of the retarded one-particle Green function, $G$, not in the new
eigen-orbital $|KJ\rangle$ basis, but in the $|kj\rangle$ basis of the normal cell instead:
\begin{equation}\label{eqn:eqn1}
G_{kj,k\prime j\prime}^{-1}(\omega) = G_{0 kj,kj}^{-1}(\omega) \delta_{k,k\prime} \delta_{j,j\prime} - V_{kj,k\prime j\prime},
\end{equation}
where $G_0$ represents a conceptual system with the period of the normal cell before $V$ is
applied.
Clearly, $G$ smoothly recovers the original period of $G_0$ as $V$ approaches zero.
Thus,
\begin{eqnarray}\label{eqn:eqn2}
 A_{kj,kj}(\omega) = \sum_{KJ}|\langle kj|KJ\rangle|^2 A_{KJ,KJ}(\omega)
\end{eqnarray}
should resemble the band structure of the normal cell with deviations in both the dispersion and
in the spectral weight that reflect the effects of $V$.
Note that while the coupling $V$ introduces non-diagonal elements of $A_{kj,k\prime j\prime}(\omega)$, we focus only on the diagonal elements here for simplicity, without loss of generality.
It is straightforward to show that in the case of $V=0$, the weight of the bands 
that follow the bands of the normal cell is exactly one, and that of the rest of the folded bands
vanishes.
One thus recovers exactly the original band structure of the normal cell as expected.
That is, \textit{the unfolded band structure is invariant against any arbitrary choice of super
cell.}

In addition, it is often desirable to also measure in each band the contribution of local
orbitals with well-defined characters (e.g.: $p_x$, $d_{xz}$, $e_g$, or bonding/anti-bonding).
This can be achieved rigorously via the use of local Wannier orbitals $|rn\rangle$:
\begin{equation}\label{eqn:eqn3}
A_{kn,kn}(\omega) = \sum_{KJ}|\langle kn|KJ\rangle |^2 A_{KJ,KJ}(\omega),
\end{equation}
where $|kn\rangle = \sum_{r} |rn\rangle \langle rn|kn\rangle  = \sum_{r} |rn\rangle  e^{ikr} / \sqrt{l}$ are the Fourier transform of the Wannier orbitals $|rn\rangle$ of orbital index $n$ and associated
with the lattice vector $r$.
(Here $l$ denotes the number of k-points in the NBZ.)
Given a consistent definition of the Wannier functions of the super-cell calculation that maps
$|RN\rangle$ of the super cell to $|R+r,n\prime\rangle$ of the normal cell,
where $r=r(N)$ is a normal-cell lattice vector within the first super cell, and
$n\prime=n\prime(N)$ is the corresponding normal-cell orbital index,
the use of Wannier function also reduces dramatically the computational expense by turning the
factor
\begin{eqnarray}\label{eqn:eqn4}
\langle kn|KJ\rangle  &=& \sum_{RN} \langle kn|RN\rangle \langle RN|KN\rangle
\langle KN|KJ\rangle \nonumber\\
&=& \sum_{RN} \langle kn|R+r(N),n\prime(N)\rangle \langle RN|KN\rangle
\langle KN|KJ\rangle  \nonumber \\
&=& \sqrt{1/Ll}\sum_{RN} e^{i(K-k)\cdot R} e^{-ik\cdot r(N)} \delta_{n,n\prime(N)}
\langle KN|KJ\rangle  \nonumber  \\
&=& \sqrt{L/l}\sum_{N} e^{-ik\cdot r(N)} \delta_{n,n\prime(N)}\delta_{[k],K} \langle KN|KJ\rangle
\end{eqnarray}
into merely a structure factor that is a sum of coefficients of the eigen-orbital $|KJ\rangle $
of the super cell in the Wannier function basis, modulated by the proper phase that encapsulates
the internal position in the super cell.
Here $[k]$ denotes the k-point folded into the SBZ from $k$.
Since $A_{KJ,KJ}$ is just a delta function at the eigenvalue $\delta(\omega-\epsilon_{KJ})$,
this final expression in essence requires only a simple coding to plot all the eigenvalues of the
super cell in the larger NBZ with a proper weight.

Of course, the above definition only makes sense when the Wannier functions
$|RN\rangle \leftrightarrow |rn\rangle $ and
$|RN \prime\rangle \leftrightarrow |r\prime n\rangle $ (that are translational symmetric in the
normal cell unit: same $n$ different $r$) are approximately identical.
Therefore, the ``gauge''~\cite{Vanderbilt} of constructing $|RN\rangle $ and $|RN\prime\rangle$ 
\textit{with the same $n$} must be controlled accordingly.
In the presence of a potential that breaks the translational symmetry of the normal cell, for
example, coming from a CDW, lattice distortions, impurities, etc., the commonly
employed~\cite{Thygesen,Wang,Eiguren} maximally localized Wannier function~\cite{Vanderbilt}
and other minimization-based methods~\cite{Gygi,Giustino} risk defining the gauge differently in the
super cell in favor of better localization, and thus should be used with extreme caution.
We found that a maximum projection method~\cite{Andersen, Ku, Anisimov} with consistent projection
between the normal and the super cells works well to satisfy this requirement.
Equations (\ref{eqn:eqn3}) and (\ref{eqn:eqn4}) should in principle also be applicable in many
existing codes employing atomic center local orbitals as basis~\cite{Koepernik, Scheffler},
as long as the non-orthogonal nature of those bases is taken into account.
Of course, these methods do not benifit from the energy resolution of the Wannier functions
that allows unfolding only the bands within the physically relevant energy range.

The unfolded band structure also has an important direct connection to the ARPES measurement.
For systems with enlarged unit cells due to weak symmetry breaking, the ARPES spectra typically
shows different band structures in different Brillouin zones of the super cell, distinctly
different from the results of first-principles calculations, which have all the bands in the SBZ.
In some cases, the observed ARPES spectra might even appear ignorant about the SBZ~\cite{Xu}.
This significant mismatch is typically regarded as the effect of the ``matrix element'' and left
unaddressed by both theorists and experimentalists, making a direct comparison very difficult.
Within the ``sudden approximation'', the ARPES intensity is proportional to~\cite{Caroli,Bansil}
\begin{eqnarray}\label{eqn:eqn5}
& & \sum_{KJ} | \textbf{e}\cdot \langle f|\textbf{p}|KJ\rangle |^2 A_{KJ,KJ}(\omega) \nonumber \\
&\sim& \sum_{KJkn} |\textbf{e}\cdot \langle f|\textbf{p}|kn\rangle |^2 | \langle kn|KJ\rangle |^2
A_{KJ,KJ}(\omega) \nonumber \\
&=& \sum_{kn} | \textbf{e}\cdot \langle f|\textbf{p}|kn\rangle |^2 A_{kn,kn}(\omega), 
\end{eqnarray}
where $\textbf{e}$ denotes the polarization vector of light, and $|f\rangle$ the ``final state''
of the photoelectron.
Clearly, except the polarization dependent dipole matrix element,
$|\textbf{e}\cdot \langle f|\textbf{p}|kn\rangle |^2$, the unfolded spectral function,
$A_{kn,kn}(\omega)$, contains almost the full information of the experimental spectrum by
absorbing the additional structure factor $\langle kn|KJ\rangle$, absent in the
typical super-cell solution, $A_{KJ,KJ}(\omega)$.
Obviously, the inclusion of this additional matrix element would facilitate significantly the
comparison between the theory and the ARPES experiment.

\begin{figure}[tbp]
\includegraphics[width=0.9\columnwidth,clip=true]{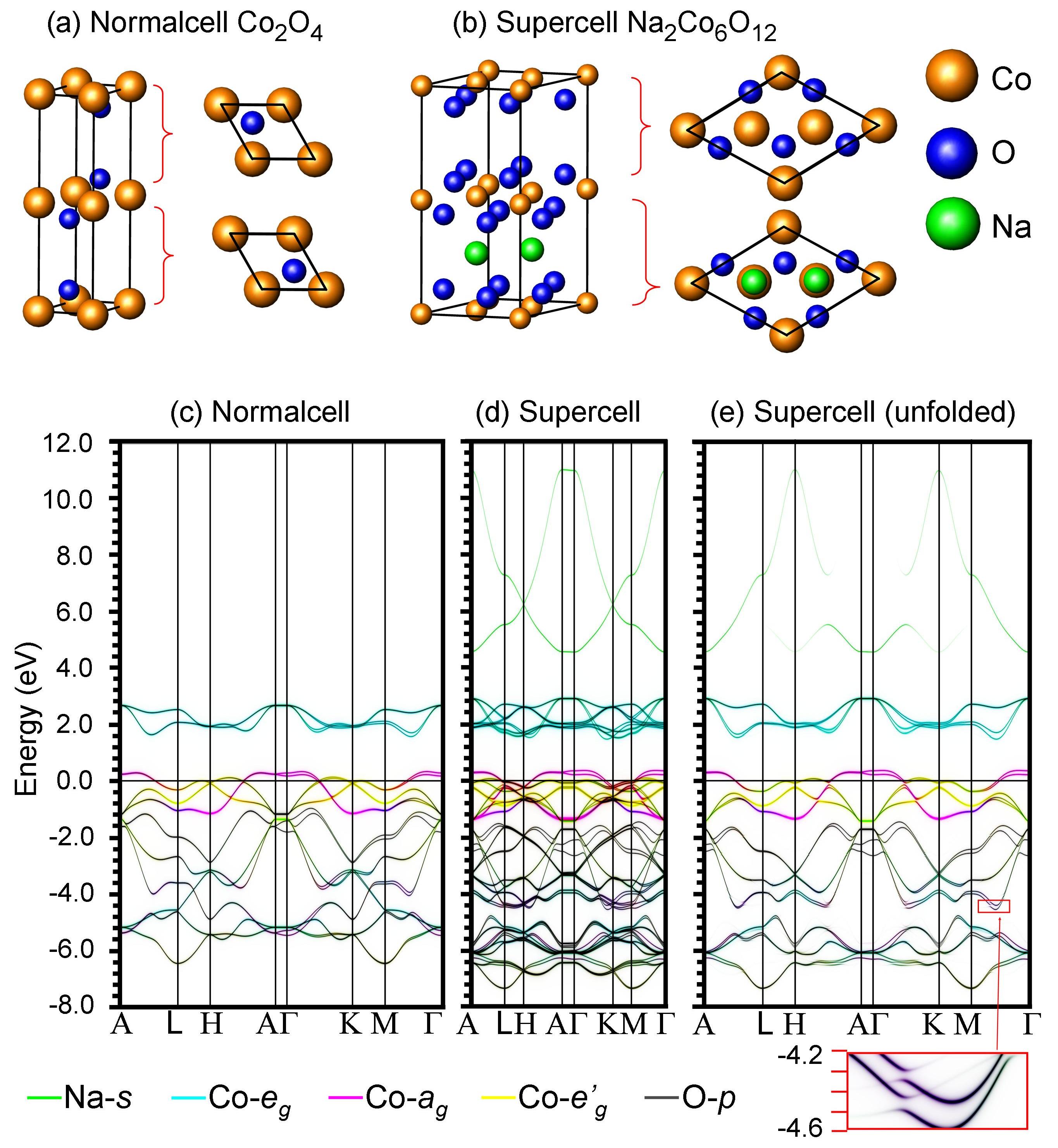}
\caption{\label{fig:fig2} (color online).
Lattice structures of (a) Co$_2$O$_4$ (normal cell) and (b) Na$_2$Co$_6$O$_{12}$ (super cell),
the corresponding band structure of (c) the normal cell and (d) super cell calculation, and
(e) the unfolded band structure of the super cell.
Insect illustrates the effects of weak translational symmetry breaking via spectral functions
over the region [-4.6eV,-4.2eV] and [$\frac{2}{5}\Gamma M$, $\frac{1}{5}\Gamma M$].
}
\end{figure}

As an example, let's consider the effect of Na impurities in Na-doped cobaltates, Na$_x$CoO$_2$
at $x=1/3$.
In typical first-principles studies\cite{Singh, Johannes}, the impurity is incorporated via a
super cell as demonstrated in Fig.~\ref{fig:fig2}(b) in comparison with the undoped normal cell
shown in Fig.~\ref{fig:fig2}(a).
Fig.~\ref{fig:fig2}(d) and (c) show the corresponding band structures obtained with standard DFT
calculations.
Since in this example the super cell is three times larger than the normal cell, the
corresponding SBZ is three times smaller and contains three times more bands.
Clearly, even for such a small super cell, the change of the size/orientation of the SBZ and
more importantly the large number of folded bands, make it practically impossible to cleanly
compare with the band structure in the NBZ of the undoped parent compound.
In fact, to many untrained eyes, these two band structures may appear entirely unrelated.

By contrast, the unfolded band structure shown in Fig.~\ref{fig:fig2}(e), demonstrates a strong
resemblance to the band structure of the undoped compound.
This allows a clear visualization of the effects of the (periodic) Na impurities on the original
Co and O bands.
Specifically, besides the introduction of additional Na-$s$ bands, one observes shifts in band
energies, gap openings and the nearby ``shadow bands'', all of which reflects the
influence of the Na impurity on these bands.
What is really nice here is the cleanness of the unfolded band structure in general, owing to the
weak intensity of the shadow bands.
As expected, the influence of the Na impurity is only minor on most Co-$d$ and O-$p$ bands, while
the Na-$s$ bands themselves show sizable effects of broken translational symmetry.
The size of the gap opening and the intensity of the shadow bands actually reflect directly the
strength of each band's coupling to the broken translational symmetry of the normal cell
(in this specific case, to the charge-density-wave order parameter introduced by the periodic
presence of Na atoms.)
Of course, for a simulation of randomly positioned impurities, these CDW-related features are
entirely artificial, and the unfolded band structure makes apparent the alarming limitation of
common practice of using small super cells in the study of impurities.
On the other hand, in many other cases, for example the super modulation of the lattice, these
features would actually correspond to a physical order parameter and provide valuable
information.

\begin{figure}[tbp]
\includegraphics[width=0.9\columnwidth,clip=true]{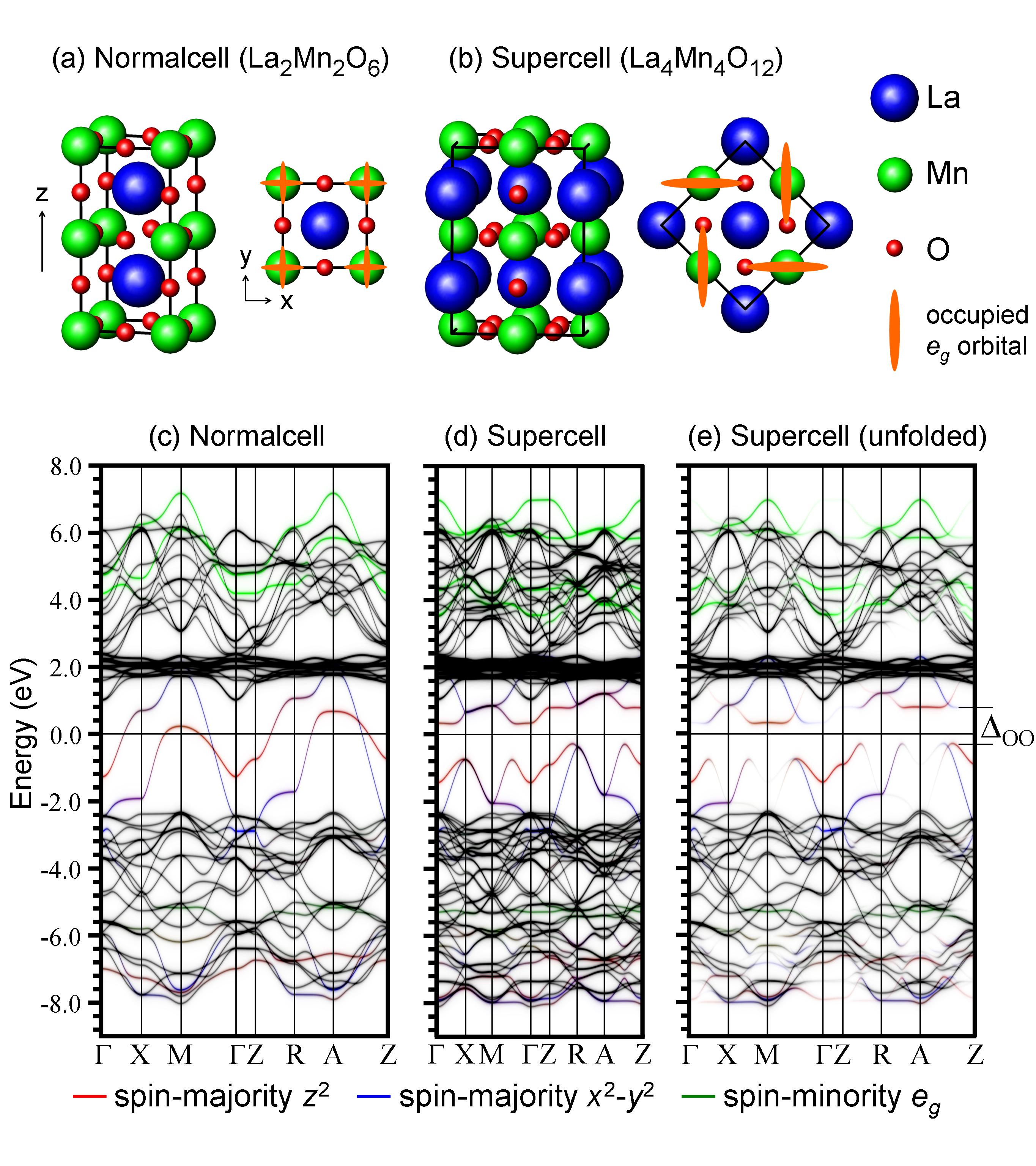}
\caption{\label{fig:fig3} (color online).
Lattice structures of (a) normal cell and (b) doubled-size supercell of A-type
anti-ferromagnetic ordered LaMnO$_3$, without and with the orbital-ordering, respectively,
the corresponding band structure of the normal cell (c) and the super cell (d), and the unfolded band structure of the super cell (e) indicating the orbital ordering gap, $\Delta_{OO}$.
Red and blue bands denote the $z^{2}$ and $x^{2}-y^{2}$ orbital characters of the spin-majority
channel, and green bands gives both $e_g$ characters of the spin-minority channel.
}
\end{figure}

As another example, let's consider a spontaneous orbital ordering in A-type anti-ferromagnetic
LaMnO$_3$.
Figure~\ref{fig:fig3} (c) and (d) show the similar comparison of band structures without and
with the long-range staggered orbital order, corresponding to unit cells shown in
Fig.~\ref{fig:fig3} (a) (normal cell) and Fig.~\ref{fig:fig3} (b) (super cell), respectively.
Both results are obtained via the LSDA+$U$ ($U$=8eV, $J_H$=0.88eV) approximation without lattice
relaxation for simplicity, without loss of generality.
By comparing the band structures with and without the orbital order on the equal footing,
the detailed information of the spontaneous orbital order should be visualized explicitly.

Just like in the Na$_x$CoO$_2$ case, the straightforward results of the orbital ordered (OO) band
structure (Fig.~\ref{fig:fig3} (d)) of the super cell calculation~\cite{Pickett} can hardly be compared with the
non-OO one (Fig.~\ref{fig:fig3} (c)).
By contrast, the unfolded band structure (Fig.~\ref{fig:fig3} (e)) of the OO case resembles 
strongly the non-OO case.
In fact, one finds that only those bands of Mn-$e_g$ character (red, blue, and green) show strong coupling to the OO order parameter with large gap openings and intensive shadow bands,
while the rest of the bands are basically uncoupled to the orbital order.
In addition, from the significant energy gain corresponding to the large OO gap ($\Delta_{OO}$)
near the Fermi level of the red and blue bands, it is apparent that essentially the orbital order
is driven only by the spin-majority $e_g$-orbitals (red and blue).
All these effects are of course entirely consistent with the existing ``electronic interaction
assisted Jahn-Teller picture''~\cite{Dagotto, Yin, Volja}, in which the degenerate Mn-$e_g$ orbitals split to
gain energy and stabilize the system at low-temperature.
However, this unfolded band structure represents probably the best visualization of such physics
in real materials with details of first-principles calculations.

In conclusion, a simple method for unfolding first-principles band structures of super cell
calculations is presented.
Proper spectral weights are obtained with negligible computational cost by making use of the
Kohn-Sham orbitals with the help of carefully chosen Wannier functions.
The inclusion of the structure factor in the resulting unfolded band structure makes it
ideal for direct comparison with the ARPES measurement.
The resulting unfolded band structures allow an easy visualization of each band's coupling to
the order parameter of spontaneous broken translational symmetry, as well as their couplings to
the external symmetry breakers like the impurities and lattice distortions.
Our method should prove valuable in the study of a wide range of problems requiring the use of
super cells, including systems with impurities, vacancies, and lattice distortions, and
broken symmetry phases of strongly correlated materials, to name a few.

This work was supported by the U.S. Department of Energy,
Office of Basic Energy Science, under Contract No. DE-AC02-98CH10886, and DOE-CMSN.

\bibliography{refs}

\begin{thebibliography}{23}
\expandafter\ifx\csname natexlab\endcsname\relax\def\natexlab#1{#1}\fi
\expandafter\ifx\csname bibnamefont\endcsname\relax
  \def\bibnamefont#1{#1}\fi
\expandafter\ifx\csname bibfnamefont\endcsname\relax
  \def\bibfnamefont#1{#1}\fi
\expandafter\ifx\csname citenamefont\endcsname\relax
  \def\citenamefont#1{#1}\fi
\expandafter\ifx\csname url\endcsname\relax
  \def\url#1{\texttt{#1}}\fi
\expandafter\ifx\csname urlprefix\endcsname\relax\def\urlprefix{URL }\fi
\providecommand{\bibinfo}[2]{#2}
\providecommand{\eprint}[2][]{\url{#2}}

\bibitem[{\citenamefont{Kohn and Sham}(1965)}]{KohnSham}
\bibinfo{author}{\bibfnamefont{W.}~\bibnamefont{Kohn}} \bibnamefont{and}
  \bibinfo{author}{\bibfnamefont{L.~J.} \bibnamefont{Sham}},
  \bibinfo{journal}{Phys. Rev.} \textbf{\bibinfo{volume}{140}},
  \bibinfo{pages}{A1133} (\bibinfo{year}{1965}).

\bibitem[{\citenamefont{Hohenberg and Kohn}(1964)}]{HohenbergKohn}
\bibinfo{author}{\bibfnamefont{P.}~\bibnamefont{Hohenberg}} \bibnamefont{and}
  \bibinfo{author}{\bibfnamefont{W.}~\bibnamefont{Kohn}},
  \bibinfo{journal}{Phys. Rev.} \textbf{\bibinfo{volume}{136}},
  \bibinfo{pages}{B864} (\bibinfo{year}{1964}).

\bibitem[{\citenamefont{Ashcroft and Mermin}(1976)}]{AshcroftMermin}
\bibinfo{author}{\bibfnamefont{N.~W.} \bibnamefont{Ashcroft}} \bibnamefont{and}
  \bibinfo{author}{\bibfnamefont{N.~D.} \bibnamefont{Mermin}},
  \emph{\bibinfo{title}{Solid State Physics}} (\bibinfo{publisher}{Holt,
  Rinehart, and Winston}, \bibinfo{address}{New York}, \bibinfo{year}{1976}).

\bibitem[{\citenamefont{Marzari and Vanderbilt}(1997)}]{Vanderbilt}
\bibinfo{author}{\bibfnamefont{N.}~\bibnamefont{Marzari}} \bibnamefont{and}
  \bibinfo{author}{\bibfnamefont{D.}~\bibnamefont{Vanderbilt}},
  \bibinfo{journal}{Phys. Rev. B} \textbf{\bibinfo{volume}{56}},
  \bibinfo{pages}{12847} (\bibinfo{year}{1997}).

\bibitem[{\citenamefont{Thygesen et~al.}(2005)\citenamefont{Thygesen, Hansen,
  and Jacobsen}}]{Thygesen}
\bibinfo{author}{\bibfnamefont{K.~S.} \bibnamefont{Thygesen}},
  \bibinfo{author}{\bibfnamefont{L.~B.} \bibnamefont{Hansen}},
  \bibnamefont{and} \bibinfo{author}{\bibfnamefont{K.~W.}
  \bibnamefont{Jacobsen}}, \bibinfo{journal}{Phys. Rev. B}
  \textbf{\bibinfo{volume}{72}}, \bibinfo{pages}{125119}
  (\bibinfo{year}{2005}).

\bibitem[{\citenamefont{Wang et~al.}(2006)\citenamefont{Wang, Yates, Souza, and
  Vanderbilt}}]{Wang}
\bibinfo{author}{\bibfnamefont{X.}~\bibnamefont{Wang}},
  \bibinfo{author}{\bibfnamefont{J.~R.} \bibnamefont{Yates}},
  \bibinfo{author}{\bibfnamefont{I.}~\bibnamefont{Souza}}, \bibnamefont{and}
  \bibinfo{author}{\bibfnamefont{D.}~\bibnamefont{Vanderbilt}},
  \bibinfo{journal}{Phys. Rev. B} \textbf{\bibinfo{volume}{74}},
  \bibinfo{pages}{195118} (\bibinfo{year}{2006}).

\bibitem[{\citenamefont{Eiguren and Ambrosch-Draxl}(2008)}]{Eiguren}
\bibinfo{author}{\bibfnamefont{A.}~\bibnamefont{Eiguren}} \bibnamefont{and}
  \bibinfo{author}{\bibfnamefont{C.}~\bibnamefont{Ambrosch-Draxl}},
  \bibinfo{journal}{Phys. Rev. B} \textbf{\bibinfo{volume}{78}},
  \bibinfo{pages}{045124} (\bibinfo{year}{2008}).

\bibitem[{\citenamefont{Gygi et~al.}(2003)\citenamefont{Gygi, Fattebert, and
  Schwegler}}]{Gygi}
\bibinfo{author}{\bibfnamefont{F.}~\bibnamefont{Gygi}},
  \bibinfo{author}{\bibfnamefont{J.-L.} \bibnamefont{Fattebert}},
  \bibnamefont{and}
  \bibinfo{author}{\bibfnamefont{E.}~\bibnamefont{Schwegler}},
  \bibinfo{journal}{Comput. Phys. Commun.} \textbf{\bibinfo{volume}{155}},
  \bibinfo{pages}{1} (\bibinfo{year}{2003}).

\bibitem[{\citenamefont{Giustino and Pasquarello}(2006)}]{Giustino}
\bibinfo{author}{\bibfnamefont{F.}~\bibnamefont{Giustino}} \bibnamefont{and}
  \bibinfo{author}{\bibfnamefont{A.}~\bibnamefont{Pasquarello}},
  \bibinfo{journal}{Phys. Rev. Lett.} \textbf{\bibinfo{volume}{96}},
  \bibinfo{pages}{216403} (\bibinfo{year}{2006}).

\bibitem[{\citenamefont{Andersen and Saha-Dasgupta}(2000)}]{Andersen}
\bibinfo{author}{\bibfnamefont{O.~K.} \bibnamefont{Andersen}} \bibnamefont{and}
  \bibinfo{author}{\bibfnamefont{T.}~\bibnamefont{Saha-Dasgupta}},
  \bibinfo{journal}{Phys. Rev. B} \textbf{\bibinfo{volume}{62}},
  \bibinfo{pages}{R16219} (\bibinfo{year}{2000}).

\bibitem[{\citenamefont{Ku et~al.}(2000)}]{Ku}
\bibinfo{author}{\bibfnamefont{W.}~\bibnamefont{Ku}} \bibnamefont{et~al.},
  \bibinfo{journal}{Phys. Rev. Lett.} \textbf{\bibinfo{volume}{89}},
  \bibinfo{pages}{167204} (\bibinfo{year}{2000}).

\bibitem[{\citenamefont{Anisimov et~al.}(2005)}]{Anisimov}
\bibinfo{author}{\bibfnamefont{V.~I.} \bibnamefont{Anisimov}}
  \bibnamefont{et~al.}, \bibinfo{journal}{Phys. Rev. B}
  \textbf{\bibinfo{volume}{71}}, \bibinfo{pages}{125119}
  (\bibinfo{year}{2005}).

\bibitem[{\citenamefont{Koepernik and Eschrig}(1999)}]{Koepernik}
\bibinfo{author}{\bibfnamefont{K.}~\bibnamefont{Koepernik}} \bibnamefont{and}
  \bibinfo{author}{\bibfnamefont{H.}~\bibnamefont{Eschrig}},
  \bibinfo{journal}{Phys. Rev. B} \textbf{\bibinfo{volume}{59}},
  \bibinfo{pages}{1743} (\bibinfo{year}{1999}).

\bibitem[{\citenamefont{Blum et~al.}(2009)}]{Scheffler}
\bibinfo{author}{\bibfnamefont{V.}~\bibnamefont{Blum}} \bibnamefont{et~al.},
  \bibinfo{journal}{Comput. Phys. Commun.} \textbf{\bibinfo{volume}{180}},
  \bibinfo{pages}{2175} (\bibinfo{year}{2009}).

\bibitem[{\citenamefont{Lee et~al.}(2010)\citenamefont{Lee, Xu, Ku, Wen, Lee,
  Katayama, Xu, Ji, Lin, Gu et~al.}}]{Xu}
\bibinfo{author}{\bibfnamefont{S.-H.} \bibnamefont{Lee}},
  \bibinfo{author}{\bibfnamefont{G.}~\bibnamefont{Xu}},
  \bibinfo{author}{\bibfnamefont{W.}~\bibnamefont{Ku}},
  \bibinfo{author}{\bibfnamefont{J.~S.} \bibnamefont{Wen}},
  \bibinfo{author}{\bibfnamefont{C.~C.} \bibnamefont{Lee}},
  \bibinfo{author}{\bibfnamefont{N.}~\bibnamefont{Katayama}},
  \bibinfo{author}{\bibfnamefont{Z.~J.} \bibnamefont{Xu}},
  \bibinfo{author}{\bibfnamefont{S.}~\bibnamefont{Ji}},
  \bibinfo{author}{\bibfnamefont{Z.~W.} \bibnamefont{Lin}},
  \bibinfo{author}{\bibfnamefont{G.~D.} \bibnamefont{Gu}},
  \bibnamefont{et~al.}, \bibinfo{journal}{arXiv:0912.3205}
  (\bibinfo{year}{2010}).

\bibitem[{\citenamefont{Caroli et~al.}(1973)\citenamefont{Caroli,
  Lederer-Rozenblatt, Roulet, and Saint-James}}]{Caroli}
\bibinfo{author}{\bibfnamefont{C.}~\bibnamefont{Caroli}},
  \bibinfo{author}{\bibfnamefont{D.}~\bibnamefont{Lederer-Rozenblatt}},
  \bibinfo{author}{\bibfnamefont{B.}~\bibnamefont{Roulet}}, \bibnamefont{and}
  \bibinfo{author}{\bibfnamefont{D.}~\bibnamefont{Saint-James}},
  \bibinfo{journal}{Phys. Rev. B} \textbf{\bibinfo{volume}{8}},
  \bibinfo{pages}{4552} (\bibinfo{year}{1973}).

\bibitem[{\citenamefont{Lindroos et~al.}(2002)\citenamefont{Lindroos,
  Sahrakorpi, and Bansil}}]{Bansil}
\bibinfo{author}{\bibfnamefont{M.}~\bibnamefont{Lindroos}},
  \bibinfo{author}{\bibfnamefont{S.}~\bibnamefont{Sahrakorpi}},
  \bibnamefont{and} \bibinfo{author}{\bibfnamefont{A.}~\bibnamefont{Bansil}},
  \bibinfo{journal}{Phys. Rev. B} \textbf{\bibinfo{volume}{65}},
  \bibinfo{pages}{054514} (\bibinfo{year}{2002}).

\bibitem[{\citenamefont{Singh and Kasinathan}(2006)}]{Singh}
\bibinfo{author}{\bibfnamefont{D.}~\bibnamefont{Singh}} \bibnamefont{and}
  \bibinfo{author}{\bibfnamefont{D.}~\bibnamefont{Kasinathan}},
  \bibinfo{journal}{Phys. Rev. Lett.} \textbf{\bibinfo{volume}{97}},
  \bibinfo{pages}{016404} (\bibinfo{year}{2006}).

\bibitem[{\citenamefont{Pillay et~al.}(2008)\citenamefont{Pillay, Johannes, and
  Mazin}}]{Johannes}
\bibinfo{author}{\bibfnamefont{D.}~\bibnamefont{Pillay}},
  \bibinfo{author}{\bibfnamefont{M.~D.} \bibnamefont{Johannes}},
  \bibnamefont{and} \bibinfo{author}{\bibfnamefont{I.~I.} \bibnamefont{Mazin}},
  \bibinfo{journal}{Phys. Rev. Lett.} \textbf{\bibinfo{volume}{101}},
  \bibinfo{pages}{246808} (\bibinfo{year}{2008}).

\bibitem[{\citenamefont{Pickett and Singh}(1996)}]{Pickett}
\bibinfo{author}{\bibfnamefont{W.~E.} \bibnamefont{Pickett}} \bibnamefont{and}
  \bibinfo{author}{\bibfnamefont{D.~J.} \bibnamefont{Singh}},
  \bibinfo{journal}{Phys. Rev. B} \textbf{\bibinfo{volume}{53}},
  \bibinfo{pages}{1146} (\bibinfo{year}{1996}).

\bibitem[{\citenamefont{Hotta et~al.}(2000)\citenamefont{Hotta, Malvezzi, and
  Dagotto}}]{Dagotto}
\bibinfo{author}{\bibfnamefont{T.}~\bibnamefont{Hotta}},
  \bibinfo{author}{\bibfnamefont{A.~L.} \bibnamefont{Malvezzi}},
  \bibnamefont{and} \bibinfo{author}{\bibfnamefont{E.}~\bibnamefont{Dagotto}},
  \bibinfo{journal}{Phys. Rev. B} \textbf{\bibinfo{volume}{62}},
  \bibinfo{pages}{9432} (\bibinfo{year}{2000}).

\bibitem[{\citenamefont{Yin et~al.}(2006)}]{Yin}
\bibinfo{author}{\bibfnamefont{W.-G.} \bibnamefont{Yin}} \bibnamefont{et~al.},
  \bibinfo{journal}{Phys. Rev. Lett.} \textbf{\bibinfo{volume}{96}},
  \bibinfo{pages}{116405} (\bibinfo{year}{2006}).

\bibitem[{\citenamefont{Volja et~al.}(2010)}]{Volja}
\bibinfo{author}{\bibfnamefont{D.}~\bibnamefont{Volja}} \bibnamefont{et~al.},
  \bibinfo{journal}{Europhys. Lett.} \textbf{\bibinfo{volume}{89}},
  \bibinfo{pages}{27008} (\bibinfo{year}{2010}).

\end{thebibliography}
\end{document}